\documentclass[%
nofootinbib,
 amsmath,amssymb,
 aps,
 prd,
twocolumn, superscriptaddress
]{revtex4-1}
\usepackage[normalem]{ulem}
\usepackage{graphicx}% Include figure files
\usepackage{dcolumn}% Align table columns on decimal point
\usepackage{bm}% bold math
\usepackage{hyperref}% add hypertext capabilities
\usepackage{footnote}
\usepackage{float}
\usepackage{xcolor}
\usepackage{physics}

\begin{document}

\title{Analytic approach to axion-like-particle emission in core-collapse supernovae}

\author{Ana Luisa {\sc Foguel}} \email{afoguel@usp.br}
 %\email{??} \\
 \affiliation{
 Instituto de Física,
Universidade de São Paulo, 05508-090 São Paulo, SP, Brazil 
}

\author{Eduardo S. {\sc Fraga}} \email{fraga@if.ufrj.br}
\affiliation{
 Instituto de F\'\i sica, Universidade Federal do Rio de Janeiro,\\
 CEP 21941-909 Rio de Janeiro, RJ, Brazil 
}

\newcommand{\al}[1]{\begin{align}\begin{aligned} #1 \end{aligned}\end{align}}
\newcommand{\GMS}[1]{\textcolor{red}{#1}}

\def\be{\begin{equation}}
\def\ee{\end{equation}}
\newcommand{\ba}{\begin{eqnarray}}
\newcommand{\ea}{\end{eqnarray}}

% useful for text
\def \ie{{\it i.e. }}
\def \eg{{\it e.g.}}
\def \etal{{\it et al.}}

% units
\def \GeV{{\, \mathrm{GeV}}}
\def \MeV{{\, \mathrm{MeV}}}

\def\ALF#1{{\color{red}[#1]}} %Ana
\def\EF#1{{\color{blue} [#1]}} %Eduardo

\begin{abstract}
We investigate the impact of a presumed axion-like-particle (ALP) emission in a core-collapse supernova explosion on neutrino luminosities and mean energies employing a relatively simple analytic description. We compute the nuclear Bremsstrahlung and Primakoff axion luminosities as functions of the protoneutron star (PNS) parameters and discuss how the ALP luminosities compete with the neutrino emission, modifying the total PNS thermal energy dissipation. Our results are publicly available in the python package \textsc{ARtiSANS}, which can be used to compute the neutrino and axion observables for different choices of parameters.
\end{abstract}

\maketitle

%\thispagestyle{title}
%\newpage
%\tableofcontents
%%%%%%%%%%%%%%%%%%%%%%%%%%%%%%%%%%%%%%%%%%%%%%%%%%%%%%%%%%%
\section{Introduction}
%%%%%%%%%%%%%%%%%%%%%%%%%%%%%%%%%%%%%%%%%%%%%%%%%%%%%%%%%%%

The Standard Model (SM) of particle physics is currently the best description of the fundamental interactions between elementary particles, with an overwhelming number of confirmed experimental predictions \cite{ParticleDataGroup:2020ssz}. Nevertheless, we know that it cannot be the final theory of Nature, and should be regarded as an effective theory \cite{Brivio:2017vri}. The reason being the existence of a plethora of different issues that the SM cannot address satisfactorily, e.g., the absence of a dark matter candidate and the evidence for nonzero neutrino masses. Such questions call for the on-going search for Beyond Standard Model (BSM) signatures.

Given the variety of possible scenarios for new physics, axion-like-particles (ALPs) have shown to be among the best candidates so far (for a recent review see e.g.~\cite{Choi:2020rgn}). Such particles arise as the pseudo Nambu-Goldstone bosons in any theory with a spontaneously broken global symmetry, which makes them a very general SM extension. Just to mention some of the best motivated ALP examples, we can cite the QCD axion~\cite{Peccei:1977hh,Weinberg:1977ma,Wilczek:1977pj,GrillidiCortona:2015jxo}, related to the breaking of the Peccei-Quinn symmetry and possible solution of the strong CP problem, the familons~\cite{Froggatt:1978nt,Davidson:1981zd,Wilczek:1982rv,Jaeckel:2013uva}, related to family symmetry breaking, and the Majoron~\cite{Chikashige:1980ui,Gelmini:1980re}, related to lepton number symmetry and that can provide a mechanism for neutrino mass generation. In addition, ALP models also present a rich phenomenology, with masses and couplings running over many orders of magnitude. Depending on the model, the ALP candidate can constitute part of the dark matter content in our Universe or even act as a dark portal to a given hidden sector~\cite{Preskill:1982cy,Dine:1982ah,Arias:2012az,Ringwald:2012hr}. Such interesting properties put them in the focus of several present and future experimental programs~\cite{Graham:2015ouw,Irastorza:2018dyq,Sikivie:2020zpn,Bertuzzo:2022fcm}.

Besides collider searches, another way to probe ALP models is via core-collapse supernovae (SNe) explosions~\cite{Lee:2018lcj,Carenza:2019pxu,Fischer:2021jfm,Calore:2021klc,Mori:2022kkh}. These events are among the most extreme and powerful astrophysical phenomena in the Universe, making them perfect laboratories to test new physics. During the explosion, one would expect a significant emission of light ALPs as a result of the dominant axion-nucleon-nucleon Bremsstrahlung interaction~\cite{Raffelt:2006cw} and the Primakoff process~\cite{PhysRevD.37.1237}. Therefore, the ALP emission would compete with neutrinos in the dissipation of the SN binding energy, such that, by measuring the neutrino luminosities, one could impose constraints on the ALP model couplings. In this context,  supernova SN1987A neutrino detection was of great importance for axion physics, and the SN neutrino burst measurement was able to put strong constraints on both the nucleon- and photon-axion couplings~\cite{PhysRevD.42.3297,Raffelt:1987yt,Payez:2014xsa,Lee:2018lcj,Lucente:2020whw}. Considering that, in the last few decades, we had several detector upgrades and new neutrino experiments (see Ref.~\cite{Scholberg:2012id} for a nice review and Refs. \cite{Foguel:2020fjx,Mirizzi:2015eza,Olsen:2022pkn,DUNE:2020zfm} for novel neutrino detection techniques), we expect that a future SN neutrino detection would be crucial not only to shed some light onto the explosion mechanism and SN neutrino physics~\cite{Horiuchi:2018ofe,Koshio:2022zip} but also for the search of new physics.

In this paper we analyze the impact of a presumed supernova ALP emission on the neutrino luminosities and mean energies. In contrast with the usual methods, which rely on computationally expensive and complex SN numerical simulations~\cite{Fischer:2016cyd,Carenza:2019pxu,Fischer:2021jfm}, we provide a relatively simple analytic description. On the one hand, the analytic computation has the benefit of being more transparent to interpretation, besides depending on few free parameters. On the other hand, due to its simplicity, it can miss specific features that only a complete simulation can provide. Nevertheless, the analytic evaluation can still reproduce the main properties of axion and neutrino emissivities, as we shall show.

Regarding the axion BSM extension, we consider a generic ALP model where the axion-like-particles couple with nucleons and photons via the following Lagrangian:
\be
\mathcal{L}_{\rm ALP}= \frac{g_{aNN}}{2 m_N} (\bar \psi_N \gamma_\mu \gamma_5 \psi_N) \partial^\mu a -\frac{1}{4} g_{a \gamma \gamma} F_{\mu \nu} \tilde F^{\mu \nu} a \,,
\ee
where $g_{aNN}$ is the axion-nucleon coupling, $m_N$ is the nucleon mass, $\psi_N$ represent the nucleon Dirac field, $a$ represents the pseudo-scalar ALP field, $g_{a \gamma \gamma}$ is a $(\rm{energy})^{-1}$ dimension coupling that parametrizes the axion-photon interaction and $F_{\mu \nu}$ is the electromagnetic field strength tensor. The first term in this Lagrangian describes the nuclear Bremsstrahlung axion interaction, which is the dominant emission process in the SN explosion. The second term gives rise to the Primakoff process, i.e., the axion-photon conversion that occurs in the presence of external electromagnetic fields.

For the case of the SN neutrino light curves, we follow the analytic calculations described in Ref.~\cite{Suwa:2020nee}, where the authors considered a Lane--Emden solution with polytropic index $n=1$ for the description of the protoneutron star (PNS) equation of state, and also applied the diffusion approximation for the neutrino transport equations. 
We compute the nuclear Bremsstrahlung and Primakoff axion luminosities as functions of the PNS parameters to evaluate the impact of the axion emission on several variables, such as the neutrino luminosities and mean energies. The ALP luminosities compete with the neutrino emission, modifying the total PNS thermal energy rate.

Finally, we also provide an user-friendly Python package for the analytic calculation of the neutrino and axion luminosities and mean energies in the presence of ALP nuclear Bremsstrahlung and/or Primakoff emission. The \textsc{ARtiSANS} (\textbf{A}nalytic \textbf{R}e-evalua\textbf{ti}on of \textbf{S}upernova \textbf{A}xion and \textbf{N}eutrino \textbf{S}treaming) code is publicly available at \url{https://github.com/anafoguel/ARtiSANS}, where one can also find a Jupyter Notebook tutorial for usage information. 

This paper is organized as follows. In the next section, we explain in detail the analytic calculation of the nuclear Bremsstrahlung and Primakoff axion luminosities as a function of the PNS parameters. These luminosities will compete with the neutrino emission, modifying the total PNS thermal energy rate. In section~\ref{sec:results}, we show our results on the impact of the axion emission on several variables, such as the neutrino luminosities and mean energies, for different choices of ALP nuclear and photon couplings. In section~\ref{sec:validity}, we discuss the validity of the analytic approximation. We conclude in section~\ref{sec:outlook} with our final remarks and outlook.

%%%%%%%%%%%%%%%%%%%%%%%%%%%%%%%%%%%%%%%%%%%%%%%%%%%%%%%%%%%
\section{Analytic setup} \label{sec:analytic}
%%%%%%%%%%%%%%%%%%%%%%%%%%%%%%%%%%%%%%%%%%%%%%%%%%%%%%%%%%%
Following Ref.~\cite{Suwa:2020nee}, we can express the one-flavor neutrino luminosity as
\begin{eqnarray} \label{eq:nulum}
L_\nu &\approx& 1.2 \times 10^{50} \, \qty(\frac{M_{\rm PNS}}{1.4 M_{\odot}})^{4/5}
\qty(\frac{R_{\rm PNS}}{10 {\rm km}})^{-6/5}  \qty(\frac{g \beta}{3})^{-4/5} 
\nonumber\\
&&\times \qty(\frac{s}{1 k_{B} {\rm baryon}^{-1}})^{4/5}  \, {\rm erg} \, {\rm s}^{-1}  \,,
\end{eqnarray}
where $M_{\rm PNS}$ and $R_{\rm PNS}$ are the protoneutron star (PNS) mass and radius, $g$ is a parameter that accounts for the deviation from the solution of the Lane--Emden equation of state with polytropic index $n=1$, $\beta$ is a boosting factor necessary to include the effects of heavy nuclei that amplify the neutrino cross-section due to coherent scattering, $s$ is the entropy per nucleon, and $k_B$ is the Boltzmann constant. As usual, we normalize our stellar masses using the solar mass $M_\odot$, and use typical values for the mass and radius of the protoneutron star. 

The total PNS thermal energy can be written as~\cite{Suwa:2020nee}
\begin{eqnarray}
E_{\rm th} &=& 2.5 \times 10^{52} \, \qty(\frac{M_{\rm PNS}}{1.4 M_{\odot}})^{5/3} \qty(\frac{R_{\rm PNS}}{10 {\rm km}})^{-2}
\nonumber \\
&&\times \qty(\frac{s}{1 k_{B} {\rm baryon}^{-1}})^{2} \, {\rm erg} \,.
\end{eqnarray}
Given that the neutrinos carry away the PNS thermal energy, we can relate these two quantities by 
\be
\frac{d E_{\rm th} }{ d t} = - 6 L_\nu \,,
\ee
where the factor of $6$ comes from the number of neutrino and anti-neutrino flavors. Now, if we include the ALP emission, the above equation is modified to
\be \label{eq:difeq}
\frac{d E_{\rm th} }{ d t} = - 6 L_\nu - L_a \,,
\ee
where $L_a$ is the total axion luminosity, which includes the axion emission due to nucleon-nucleon Bremsstrahlung, $L_a^{\rm brem}$, and to the Primakoff process, $L_a^{\rm prim}$. In what follows, we derive the analytic expressions for these two contributions. 

%%%%%%%%%%%%%%%%%%%%%%%%
\subsection{Nuclear Bremsstrahlung}

The total nuclear Bremsstrahlung energy-loss rate per unit mass can be expressed as~\cite{Raffelt:1996wa}
\begin{eqnarray}
\epsilon_{a}^{\rm brem} &=& \alpha_{aNN} \, 1.69 \times 10^{35} \, \qty(\frac{\rho}{10^{15}\, {\rm g \, cm}^{-3}}) 
\nonumber\\
&&\times \qty(\frac{T}{\MeV})^{7/2} \, {\rm erg} \, {\rm g}^{-1}\, {\rm s}^{-1} \,,
\end{eqnarray}
where $ \alpha_{aNN} = g_{aNN}^2/4 \pi$ is the axion-nucleon fine-structure constant, $T$ is the PNS temperature and $\rho$ is the density, which follows the Lane--Emden equation of state with $n=1$~\cite{Suwa:2020nee}
\be \label{eq:rho}
\rho = \frac{M_{\rm PNS}}{4 \pi^2}  \qty(\frac{\pi}{R_{\rm PNS}})^3 \frac{\sin \xi}{\xi}\, ,
\ee
where we defined the dimensionless radius 
\be
\xi \equiv \frac{r}{R_{\rm PNS}} \pi \, ,
\ee 
with radial coordinate $r$.

Using that we can also write the temperature as a function of the PNS mass, total radius, entropy and dimensionless radial parameter $\xi$ as \cite{Suwa:2020nee} 
\begin{eqnarray} \label{eq:temp}
T &=& 30 \, \qty(\frac{M_{\rm PNS}}{1.4 M_{\odot}})^{2/3} \qty(\frac{R_{\rm PNS}}{10 {\rm km}})^{-2}  \qty(\frac{s}{1 k_{B} {\rm baryon}^{-1}}) \nonumber\\
&& \times \qty(\frac{\sin \xi }{\xi})^{2/3} \MeV \,,
\end{eqnarray}
we can obtain the axion luminosity $L_a^{\rm brem}$ by integrating the energy-loss rate per unit volume $\epsilon_{a}^{\rm brem} \times \rho$ through the PNS radial coordinate. Assuming spherical symmetry:
\be
L_a^{\rm brem} = 4 \pi \int_0^{R_{\rm PNS}} \epsilon_{a}^{\rm brem}(r)  \rho(r) \, r^2 dr \,,
\ee
and the integration gives
\begin{eqnarray} \label{eq:axBlum}
L_a^{\rm brem} &=& \alpha_{aNN} \, 2.93 \times 10^{73} \, \qty(\frac{M_{\rm PNS}}{1.4 M_{\odot}})^{13/3} \qty(\frac{R_{\rm PNS}}{10 {\rm km}})^{-10}
\nonumber\\
&&\times \qty(\frac{s}{1 k_{B} {\rm baryon}^{-1}})  \, {\rm erg} \, {\rm s}^{-1}  \,.
\end{eqnarray}
%

%%%%%%%%%%%%%%%%%%%%%%%%
\subsection{Primakoff process}
For the case of the Primakoff process, the energy loss rate per unit volume is given by~\cite{Raffelt:1996wa}
\be \label{eq:Qprim}
Q_{P} = \frac{g_{a \gamma \gamma}^2 T^7}{4 \pi} F(\kappa^2) \,,
\ee
where $T$ is the temperature, $\kappa = k_S/2T$ with $k_S$ being the screening scale, and
\be \label{eq:Fprim}
F(\kappa^2) = \frac{\kappa^2}{2 \pi^2} \int_0^\infty dx \qty[(x^2 + \kappa^2) \ln{\qty(1 + \frac{x^2}{\kappa^2})} - x^2] \frac{x}{e^x -1} \,,
\ee
where $x$ is the ratio between the photon frequency and the temperature, i.e., $x \equiv \omega/T$. 

For a non-degenerate medium, we can obtain the screening scale via the Debye--Huckel formula
\be 
\begin{aligned}
k_S^2 &= \frac{4 \pi \alpha_{\rm EM}}{T} n_N (Y_e + \sum_j Z_j^2 Y_j) \, ,
\end{aligned}
\ee
where $n_N = \rho/m_N$ is the nucleon number density, $Y_e$ is the electron fraction per baryon, and the sum goes over the different nuclear species $j$, with fraction $Y_j$, present in the medium. Since the electrons are highly degenerate, their phase-space is Pauli blocked, which implies that their contribution to the screening scale is negligible. Considering the non-degenerate regime, we can approximate the screening length to~\cite{Payez:2014xsa}
\be 
\begin{aligned}
k_S^2 & \simeq \frac{4 \pi \alpha_{\rm EM}}{T} n_N Y_p = \frac{4 \pi \alpha_{\rm EM}}{T} n_p \, ,
\end{aligned}
\ee
where $Y_p$ is the proton fraction per baryon. However, since protons are partially degenerate we expect that the effective number of targets is reduced, so that we need to consider $n_p^{\rm eff}$. Although this number can vary depending on the PNS radius and time after bounce, following Ref.~\cite{Lee:2018lcj}, here we assume that $n_p^{\rm eff} = n_p/2$ for simplicity. Hence, we can write~ 
\be 
\begin{aligned}
k_S^2 & \simeq \frac{4 \pi \alpha_{\rm EM}}{T} \frac{\rho}{m} \frac{Y_p}{2} \,.
\end{aligned}
\ee
Now, if we employ the formulas for the density and temperature as functions of the PNS mass, total radius, entropy, and radial parameter, given by eq.~\eqref{eq:rho} and eq.~\eqref{eq:temp}, respectively, we can use equations~\eqref{eq:Qprim} and~\eqref{eq:Fprim} to compute the axion luminosity due to the Primakoff process by integrating through the PNS radial component
\be \label{eq:axPlum}
L_a^{\rm prim} =\int_0^{R_{\rm PNS}} Q_P(r) \, 4 \pi r^2 dr \,.
\ee
%

%%%%%%%%%%%%%%%%%%%%%%%%%%%%%%%%%%%%%%%%%%%%%%%%%%%%%%%%%%%
\section{Results}\label{sec:results}
%%%%%%%%%%%%%%%%%%%%%%%%%%%%%%%%%%%%%%%%%%%%%%%%%%%%%%%%%%%

According to equation~\eqref{eq:difeq}, the inclusion of the nuclear Bremsstrahlung and Primakoff axion emission processes can impact the PNS energy dissipation. Hence we can numerically solve this differential equation to obtain the entropy as a function of time $s(t)$ provided that we fix an initial condition, the PNS parameters and dark axion couplings (for details on the computation, see appendix~\ref{app:difeq}). Since coherent scattering is suppressed for earlier times, we will consider an initial and a final phase in the PNS evolution, with different boosting factors $\beta_i$ and $\beta_f$ and total emitted neutrino energies $E_{\rm tot}^i$ and $E_{\rm tot}^f$. For the results and plots presented in this section we always fix the PNS parameters to $M_{\rm PNS}= 1.5 \, M_{\odot}$, $R_{\rm PNS}= 12 \, \rm{km}$, $g= 0.04$, $\beta_{\rm i} = 3$, $\beta_{\rm f} = 40$, $E_{\rm tot}^i = 4 \times 10^{52} \, {\rm erg}$ and $E_{\rm tot}^f = 10^{53} \, {\rm erg} $. We also consider a proton fraction of $Y_p = 0.3$~\cite{Fischer:2021jfm}, which corresponds to an intermediate scenario between very neutron-rich matter ($Y_p \sim 0.05$) and symmetric nuclear matter ($Y_p = 0.5$)~\cite{Menezes:2021jmw}.  Let us emphasize that all these parameters can be easily changed in the \textsc{ARtiSANS} code.

\begin{figure}[t!]
%\begin{center}
\includegraphics[width=9cm]{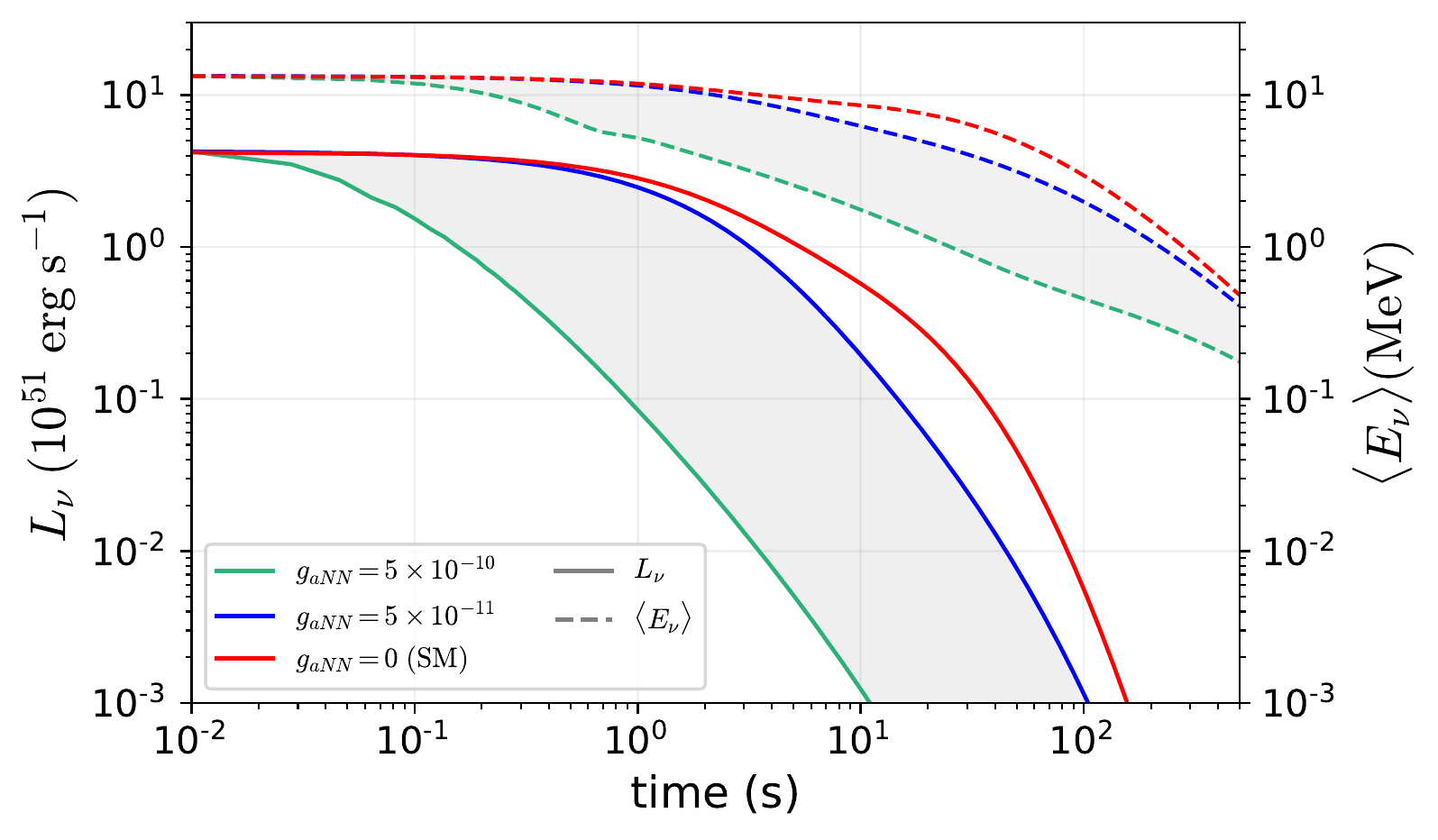}
%\end{center}
%\vglue -0.1cm
\caption{\label{fig:LnuBrem} Evolution of neutrino luminosity $L_\nu$ (solid) and mean-energy $\langle E_\nu \rangle$ (dashed) as a function of time after bounce. The red curve indicates the standard neutrino scenario while the blue (green) curve represents the case with the addition of axion emission via nuclear Bremsstrahlung for $g_{aNN} = 5 \times 10^{-11}$ ($g_{aNN} = 5 \times 10^{-10}$).}
\end{figure}

Fig.~\ref{fig:LnuBrem} shows the time evolution of the one-flavor neutrino luminosity (solid) and mean energy (dashed) for two different benchmark values of axion-nucleon coupling $g_{aNN}=5 \times 10^{-11}$ (blue), $g_{aNN}=5 \times 10^{-10}$ (green), and also for the Standard Model case where $g_{aNN}=0$ (red). For values of coupling smaller than $g_{aNN}= 10^{-11}$, the deviation of the neutrino luminosity when including axion nuclear Bremsstrahlung is less than a $\sim 5 \%$ effect.

\begin{figure}[t]
%\begin{center}
\includegraphics[width=9cm]{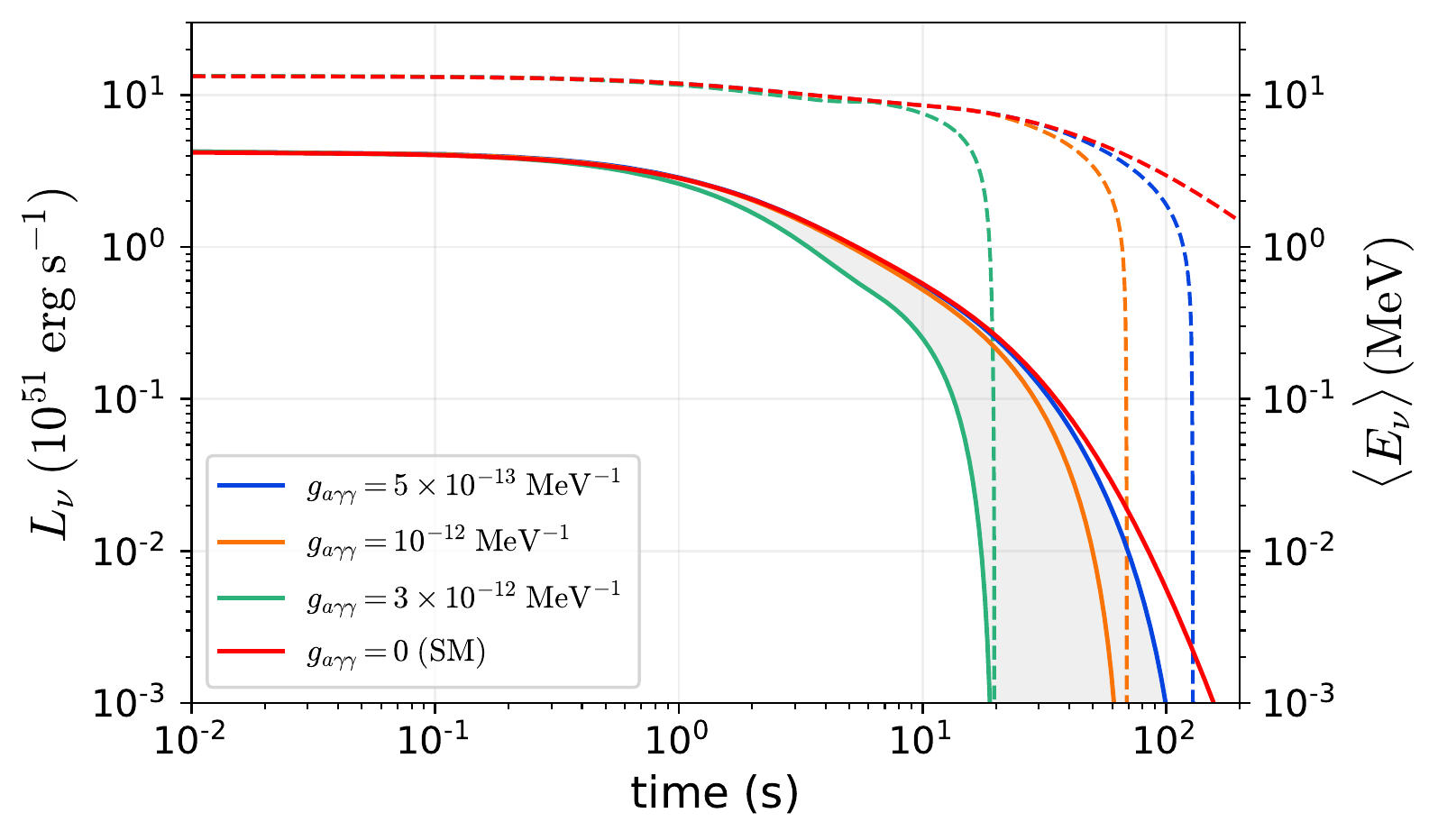}
%\end{center}
%\vglue -0.1 cm
\caption{ \label{fig:LnuPrim} Evolution of neutrino luminosity $L_\nu$ (solid) and mean-energy $\langle E_\nu \rangle$ (dashed) as a function of time after bounce. The red curve indicates the standard neutrino scenario while the blue, orange and green curves represent the case with the addition of axion emission via Primakoff effect for  the couplings $g_{a \gamma \gamma} = 5 \times 10^{-13} \, {\rm MeV}^{-1}$, $g_{a \gamma \gamma} = 10^{-12} \, {\rm MeV}^{-1}$ and $g_{a \gamma \gamma} = 3 \times 10^{-12} \, {\rm MeV}^{-1}$, respectively.}
\end{figure}

Similarly, Fig.~\ref{fig:LnuPrim} shows the time evolution of neutrino luminosities (solid) and mean energies (dashed), but when considering only Primakoff interactions. The colors indicate different choices of axion-photon couplings: $g_{a \gamma \gamma} = 5 \times 10^{-13} \, {\rm MeV}^{-1}$ (blue), $g_{a \gamma \gamma} = 10^{-12} \, {\rm MeV}^{-1}$ (orange) and $g_{a \gamma \gamma} = 3 \times 10^{-12} \, {\rm MeV}^{-1}$ (green). The red curves represent the SM case. The effect of the proton fraction is more relevant for couplings $g_{a \gamma \gamma} \gtrsim 10^{-12} \, {\rm MeV}^{-1}$, for which higher values of $Y_p$ produce larger deviations from the SM curve.
\begin{figure}[t]
%\begin{center}
\includegraphics[width=9cm]{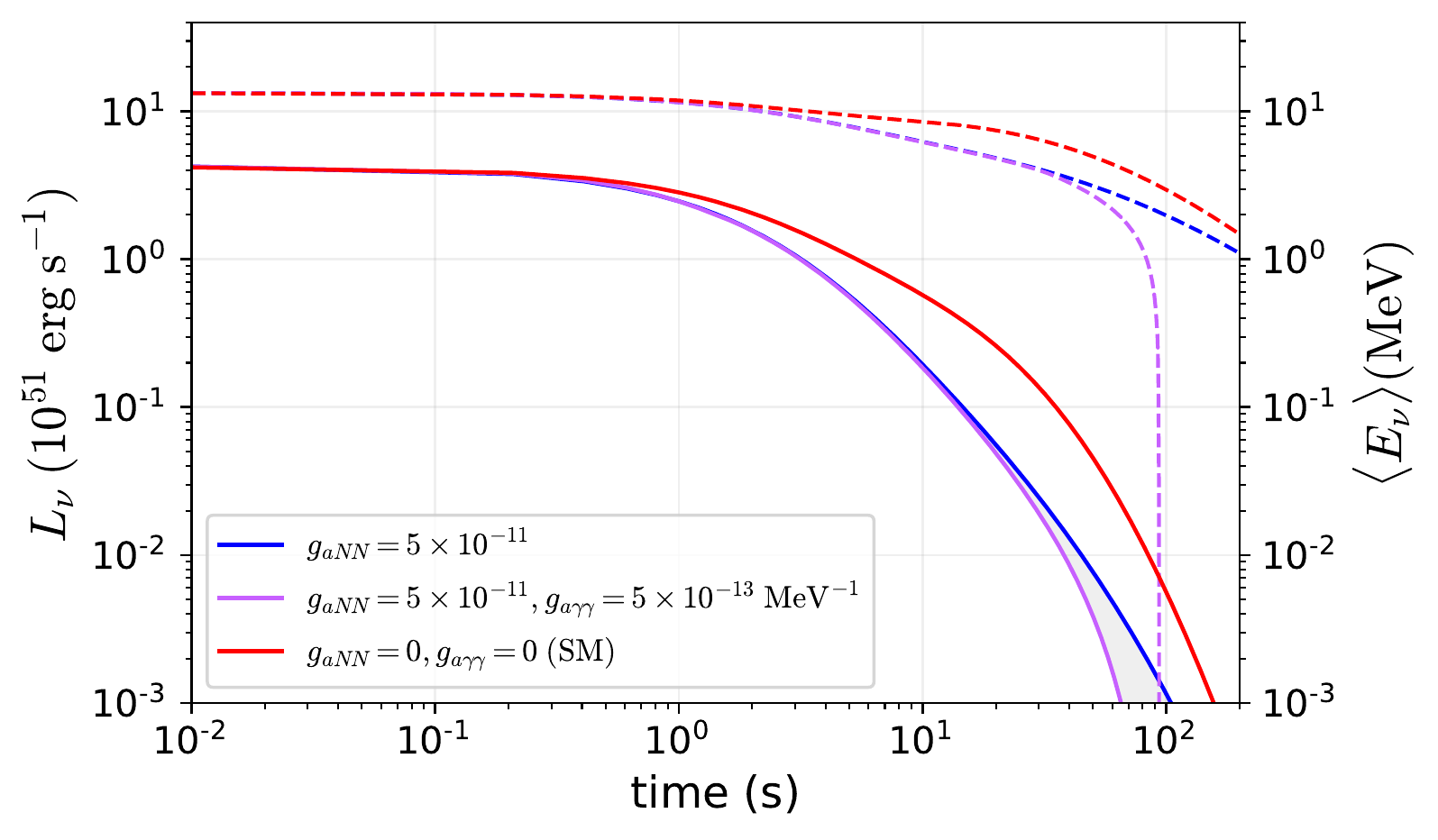}
%\end{center}
\caption{\label{fig:LnuBremPrim} Same as Fig. ~\ref{fig:LnuBrem} but also considering the effect of Primakoff emission. The blue curve is the same as before and the purple curve adds an axion-photon interaction with coupling $g_{a \gamma \gamma} = 5 \times 10^{-13} \, {\rm MeV}^{-1}$. The red curves represent the SM case with dark couplings set to zero. }
\end{figure}
In Fig.~\ref{fig:LnuBremPrim} we show the combined effect of axion nuclear Bremsstrahlung and Primakoff interactions on the neutrino variables. The blue curve is the same as in figure~\ref{fig:LnuBrem} ($g_{aNN}=5 \times 10^{-11}$) and the purple curve adds the effect of Primakoff emission with an axion-photon coupling of $g_{a \gamma \gamma} = 5 \times 10^{-13}  \, {\rm MeV}^{-1}$. We can see that the inclusion of the Primakoff interaction results in deviations on the neutrino luminosity for later times, and also has a drastic impact on the mean energies. For values of  $g_{a \gamma \gamma} \lesssim 10^{-13} \, {\rm MeV}^{-1}$ the difference from the SM case is always less than $5\%$.

\begin{figure}[t]
%\begin{center}
\includegraphics[width=8cm]{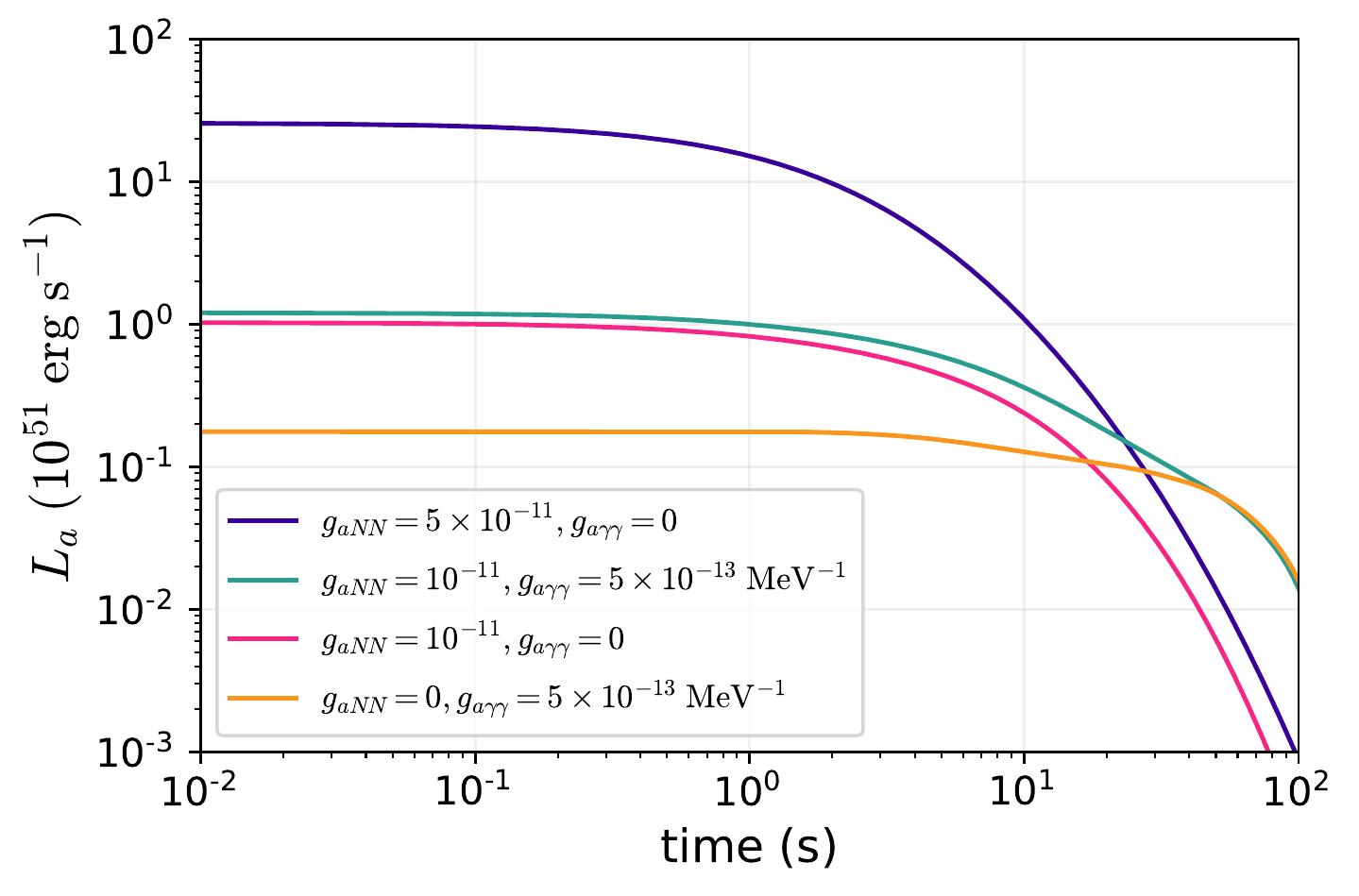}
%\end{center}
\caption{\label{fig:Laxion} Time evolution of the axion-like-particle luminosity for different combinations of dark couplings. Following the same labels as in Table~\ref{tab:totEnergy}, each color represents a specific model: MP1 (orange), MN1 (pink), MNP1 (green) and MN2 (blue).}
\end{figure}

\begin{figure}[h]
\vspace{2pt}
%\begin{center}
\includegraphics[width=0.45\textwidth]{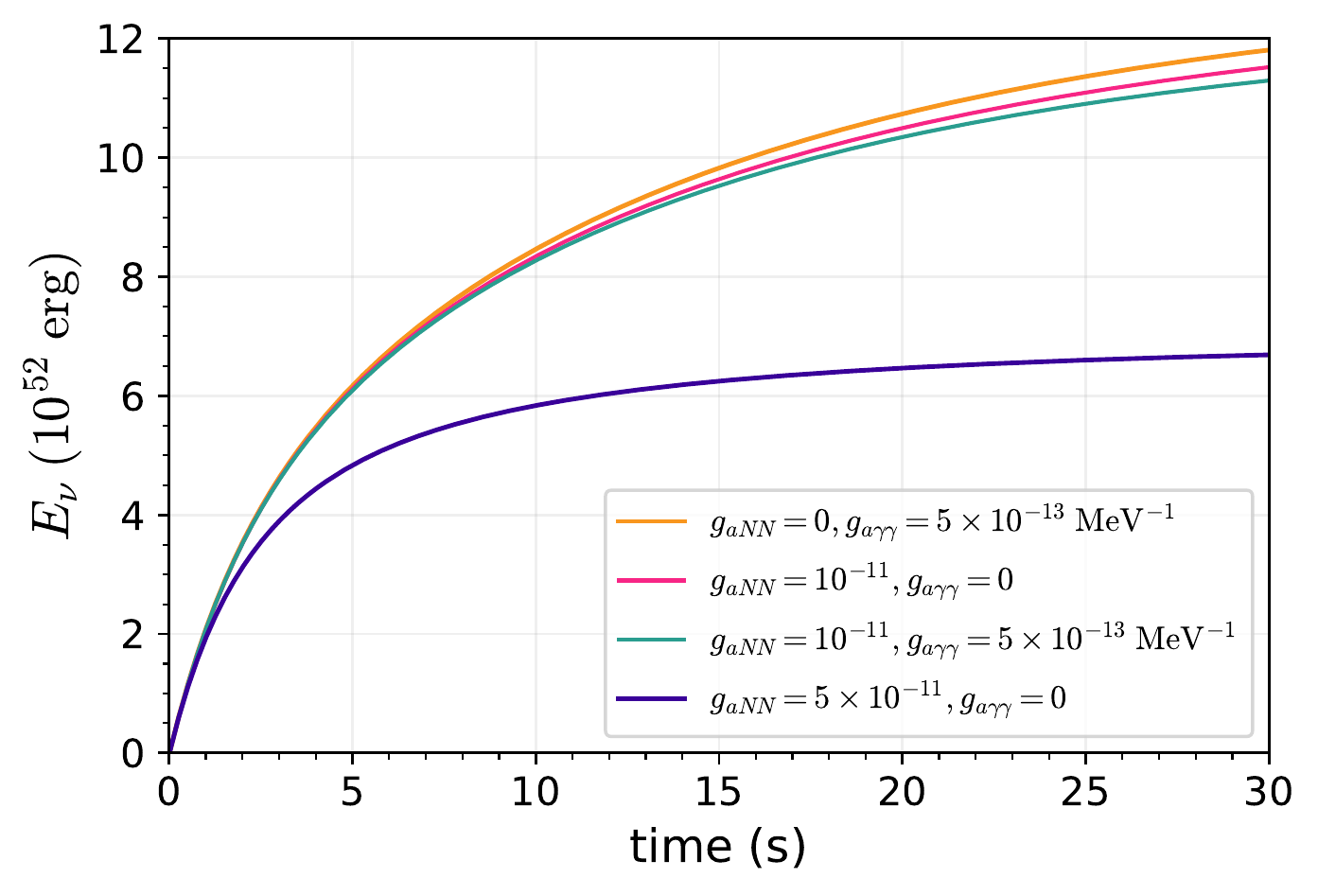}
\includegraphics[width=0.45\textwidth]{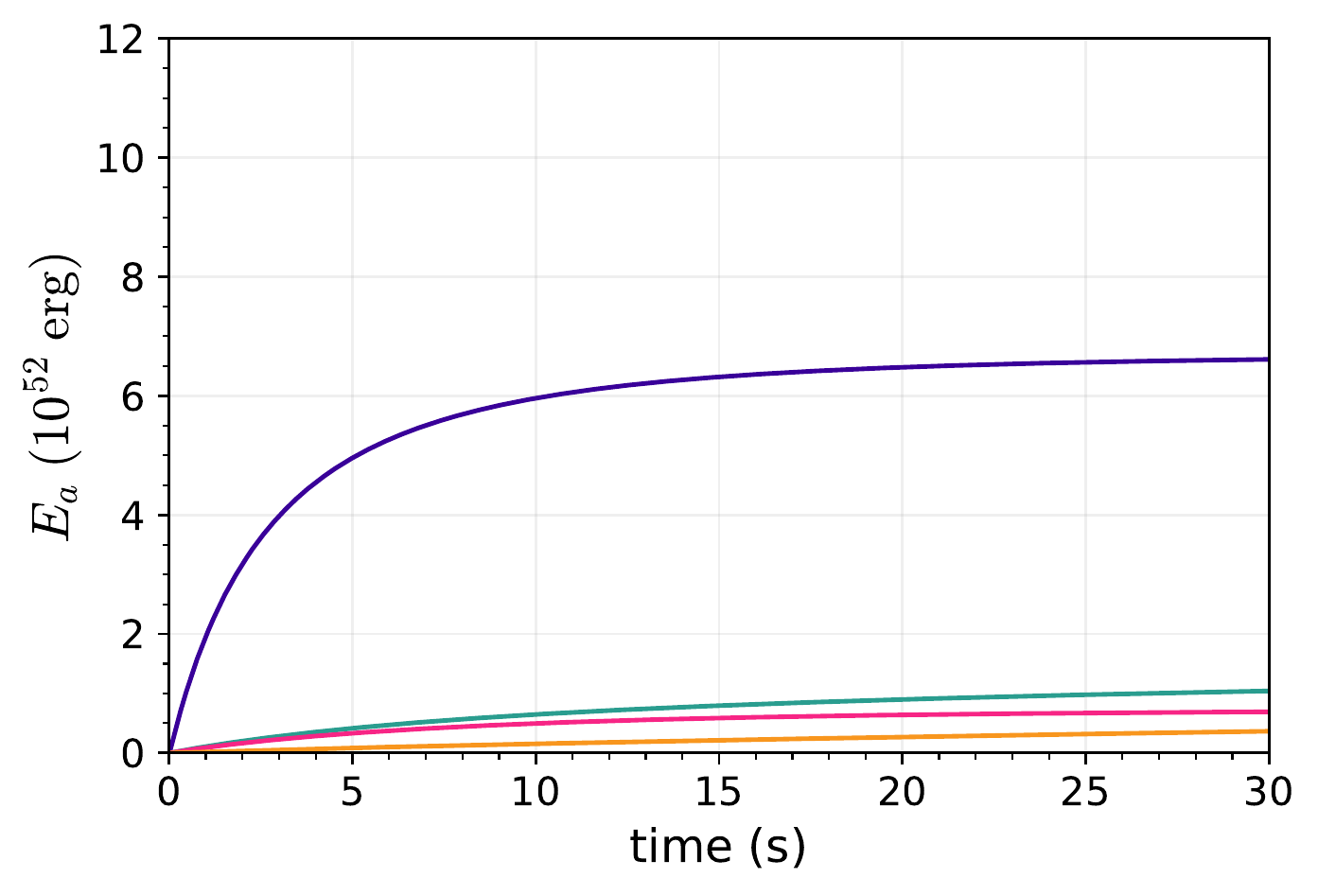}
%\end{center}
\caption{\label{fig:Etot} The upper (lower) plot shows the time evolution of the SN total energy carried away by neutrinos (axions). The colors indicate different choices of dark couplings as indicated by the labels. The neutrino emission takes into account all six flavors of neutrinos. }
\end{figure}

Concerning the luminosity released by the ALPs, we show in Fig.~\ref{fig:Laxion} the time evolution of $L_a$ for distinct benchmark values of axion-nucleon and axion-photon couplings. From the figure we can see that the nuclear Bremsstrahlung (Primakoff) axion emission is more relevant for earlier (later) times. Finally, the upper (lower) plot of Fig.~\ref{fig:Etot} shows the PNS binding energy that was carried away by neutrinos (ALPs) as a function of time. We also display in Table~\ref{tab:totEnergy} the total emitted energy by neutrinos $E_\nu$ and axions $E_a$, integrated up to 20s post-bounce, for the different choices of dark couplings, labeled by the model names in the first column.

\begin{table}[t]
\begin{center}
    \begin{tabular}{c c c c c}

         Model & $g_{aNN}$ & $g_{a \gamma \gamma} \; [ \rm{MeV}^{-1}]$ & $E_\nu \; [10^{52} \, \rm{erg}]$ & $E_a \; [10^{52} \, \rm{erg}]$  \\
        \hline 
         SM & $0$ & $0$ & $10.8$ & - \\
         MP1 & $0$ & $5 \times 10^{-13}$  &  $10.7$ &  $0.27$\\
         MN1 & $10^{-11}$ & $0$ &  $10.5$&   $0.64$\\
         MNP1 & $10^{-11}$  &  $5 \times 10^{-13}$ & $10.3$ &  $0.9$\\
         MN2 & $5 \times 10^{-11}$  & $0$ &  $6.47$ &  $6.48$\\
    \end{tabular}
    \caption{Comparison of the total energy emitted by neutrinos $E_\nu$ and axions $E_a$ for different axion parameters choices. The energies were computed by integrating the respective luminosities up to $20$s post-bounce.}
    \label{tab:totEnergy}
\end{center}  
\end{table}

%%%%%%%%%%%%%%%%%%%%%%%%%%%%%%%%%%%%%%%%%%%%%%%%%%%%%%%%%%%
\section{Validity of the Analytic Approximation}\label{sec:validity}
%%%%%%%%%%%%%%%%%%%%%%%%%%%%%%%%%%%%%%%%%%%%%%%%%%%%%%%%%%%

Although the semi-analytic method presented here can be considerably  simpler and faster than the usual computationally expensive simulations, there are still some limitations that should be kept in mind.

\begin{figure}[t]
%\begin{center}
\includegraphics[width=0.43\textwidth]{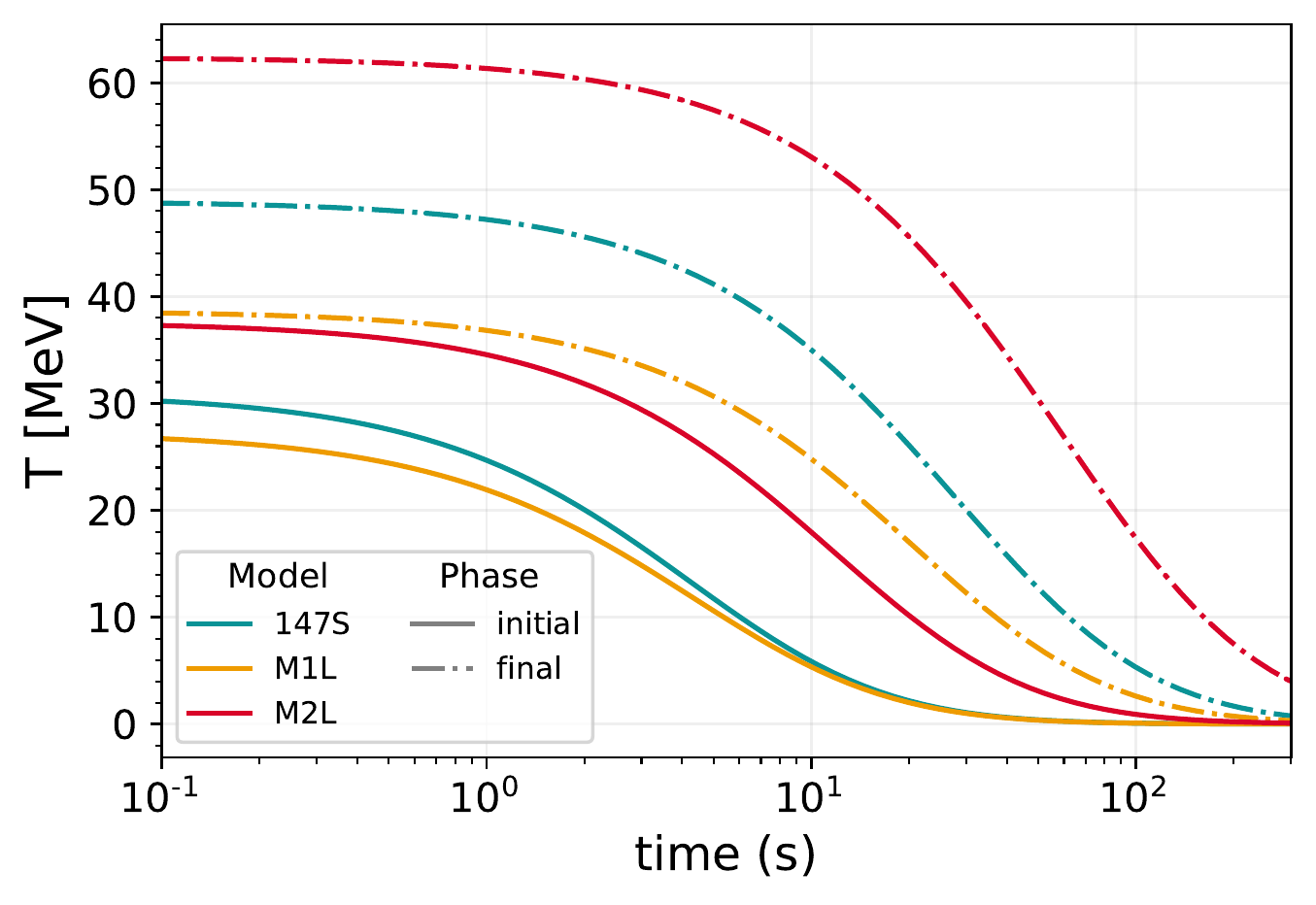}
\includegraphics[width=0.43\textwidth]{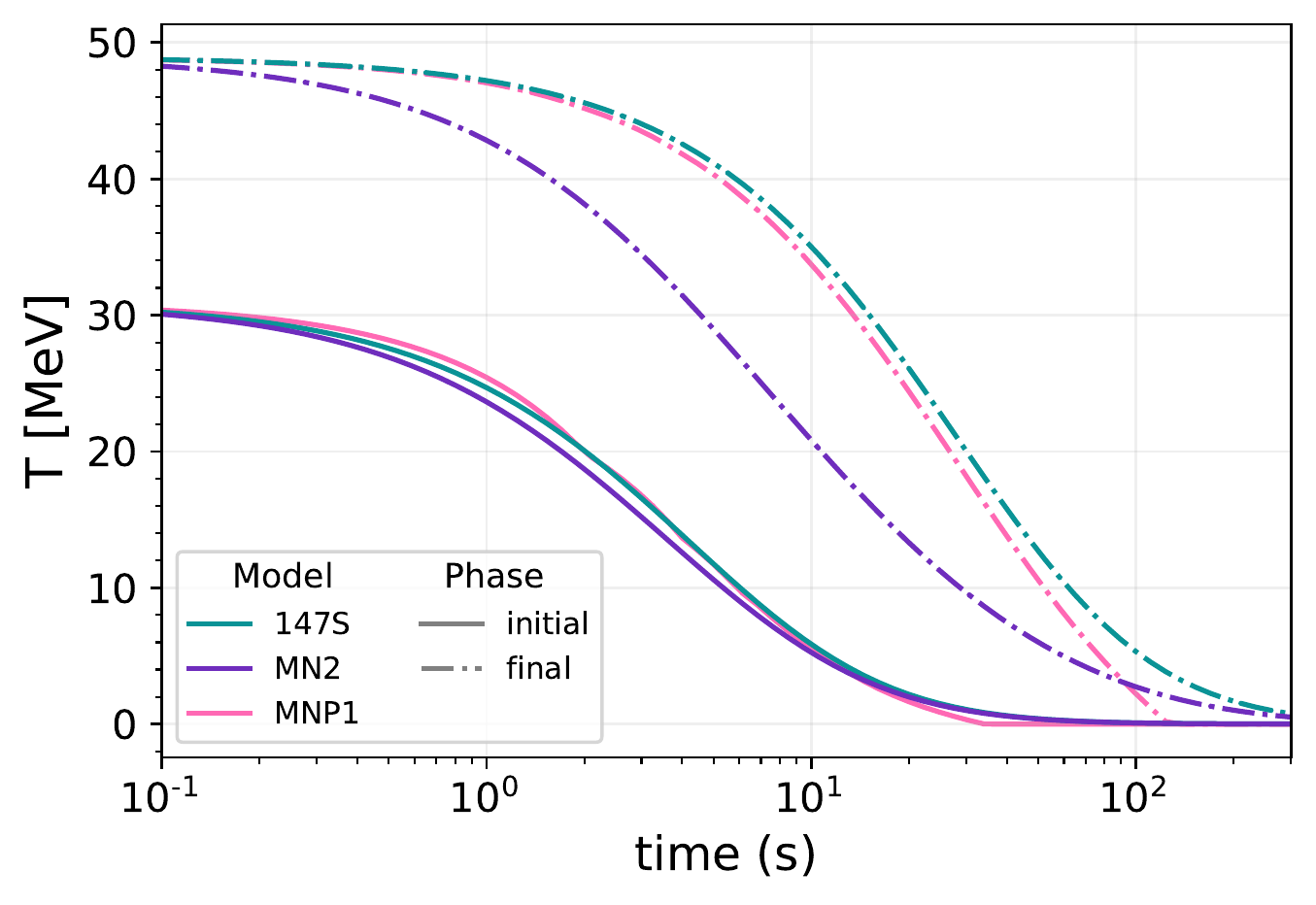}
%\end{center}
\caption{\label{fig:Temp} The upper panel shows the time evolution of the maximum temperature of models 147S (green), M1L (orange) and M2L (red) divided into the initial (solid) and final (dash-dotted) phase contributions, and computed using the analytic equation for the entropy considering only neutrinos. The lower panel shows the same, but for the axionic models MN2 (blue) and MNP1 (pink) that also include nuclear Bremsstrahlung and Primakoff processes in the entropy computation.}
\end{figure}

One important caveat that we shall discuss is related to the analytic expression for the temperature, given by equation~\eqref{eq:temp}. The upper plot of Fig.~\ref{fig:Temp} shows the maximum temperature ($\xi \rightarrow 0$) as a function of time for models 147S (green), M1L (orange) and M2L (red) of Ref.~\cite{Suwa:2020nee} (see Table~\ref{tab:PNSmodels}) separated into the initial (solid) and final (dash-dotted) phases. In this figure the temperature was computed with the expression for the entropy considering only neutrinos (equation~\eqref{eq:entropNu}). The lower panel of Fig.~\ref{fig:Temp} shows the same, but including the axionic contributions of the MN2 (blue) and MNP1 (pink) models for the entropy calculation. We can see that the overall effect of the ALP inclusion is to lower the temperatures.

\begin{table}[t]
\def\arraystretch{1.3}
\begin{center}
    \begin{tabular}{c c c c c c c c}

         Model & $M_{\rm PNS}$ & $R_{\rm PNS}$ & $g$ & $\beta_i$ &$E_{\rm tot}^i$ & $\beta_f$ &$E_{\rm tot}^f$ \\
        \hline 
         147S & $1.5 \, M_\odot$ & $12 \, {\rm km}$ & $0.04$ & $3$ & $4 \times 10^{52} \, {\rm erg}$ & $40$ & $ 10 \times10^{52} \, {\rm erg}$ \\
         M1L & $1.3 \, M_\odot$ & $11 \, {\rm km}$ & $0.04$ & $3$ & $2.5 \times 10^{52} \, {\rm erg}$ & $25$ & $ 5 \times 10^{52}\, {\rm erg}$ \\
         M2L & $2.3 \, M_\odot$ & $13 \, {\rm km}$ & $0.1$ & $3$ & $8 \times 10^{52} \, {\rm erg}$ & $30$ & $ 22 \times 10^{52} \, {\rm erg}$ \\
    \end{tabular}
    \caption{Specification of the PNS parameters used by Ref.~\cite{Suwa:2020nee} to fit the numerical solutions of the models presented in Ref.~\cite{Suwa:2019svl}.}
    \label{tab:PNSmodels}
\end{center}    
\vspace{-20pt}
\end{table}

Note that, albeit the temperatures for the initial phase always stand below $\sim 40$ MeV, final phase temperatures can be as high as $60$ MeV for initial times, depending on the model. This feature needs to be handled with caution since, according to numerical simulation~\cite{Lucente:2020whw}, temperatures cannot be bigger than $\sim 40$ MeV. In fact, in Ref.~\cite{Suwa:2020nee} the authors show that for initial times their analytic computations do not follow the numerical model 147S of Ref.~\cite{Suwa:2019svl} and, hence, we should be cautious when considering earlier times in the semi-analytic setup. This problem is actually related to the limitation of the Lane--Emden equation to model the PNS, especially for earlier times during the contraction phase. 

\begin{figure}[t]
%\begin{center}
\includegraphics[width=9cm]{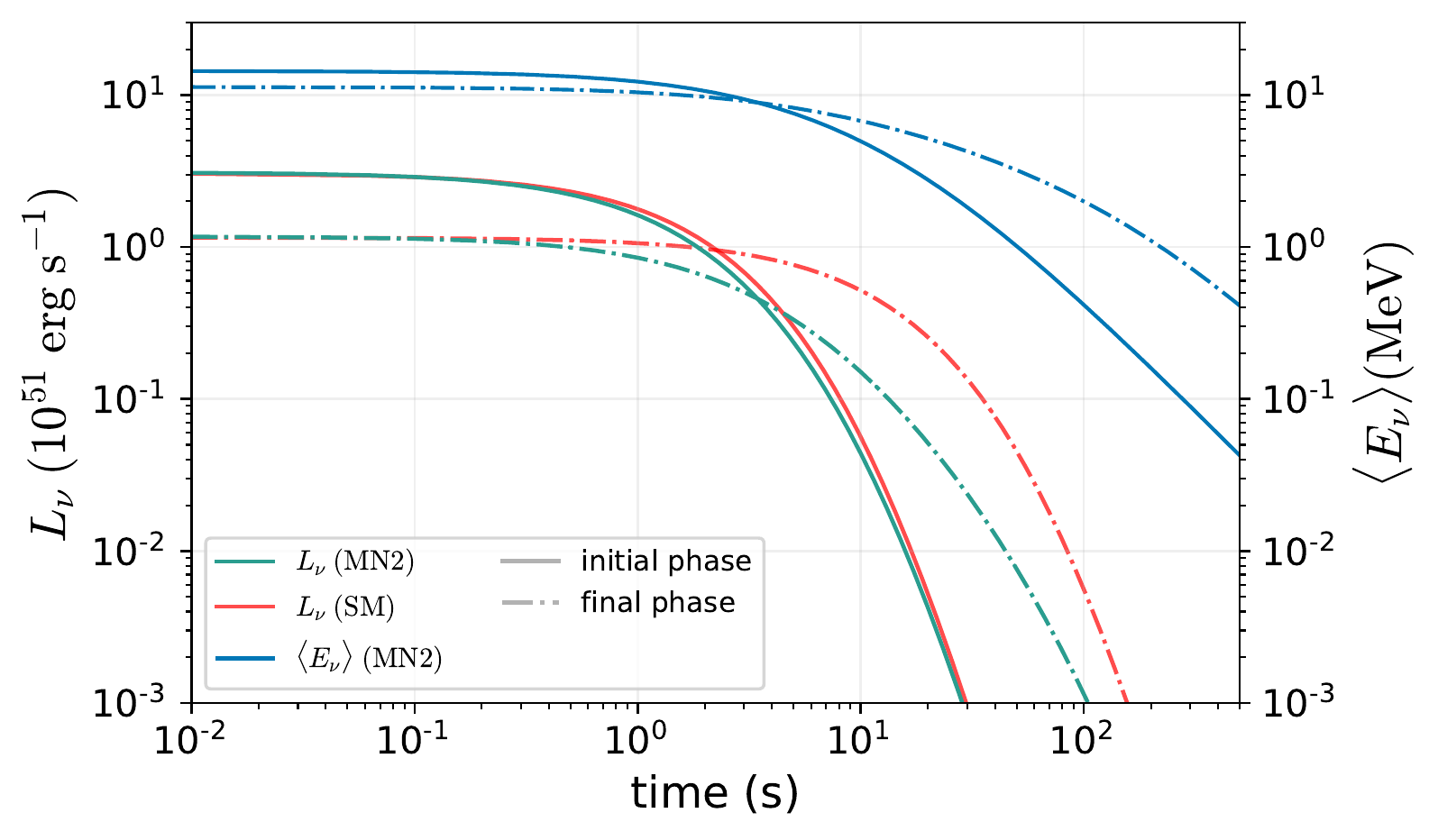}
%\end{center}
\caption{\label{fig:LnuPhases} Time evolution of the neutrino luminosity divided into initial (solid) and final (dash-dotted) phases for the MN2 model (green) and SM (red). In blue we also show the phase separation of the neutrino average energy for the MN2 model.}
\end{figure}

Let us emphasize that, for the case of neutrinos, this is a problem with the late-time phase, and this phase contribution for the neutrino luminosity is sub-dominant at earlier times. Figure~\ref{fig:LnuPhases} endorses this fact by showing the time evolution of the neutrino luminosity for the MN2 model (green) and for the SM (red) separated into initial (solid) and final (dash-dotted) phases. It also shows in blue the phase division of the neutrino average energy for the MN2 model. We can see that up to $\sim 5$ s, the initial phase is the most relevant contribution for both the luminosity and the average neutrino energy. 

\begin{figure}[t]
%\begin{center}
\includegraphics[width=8cm]{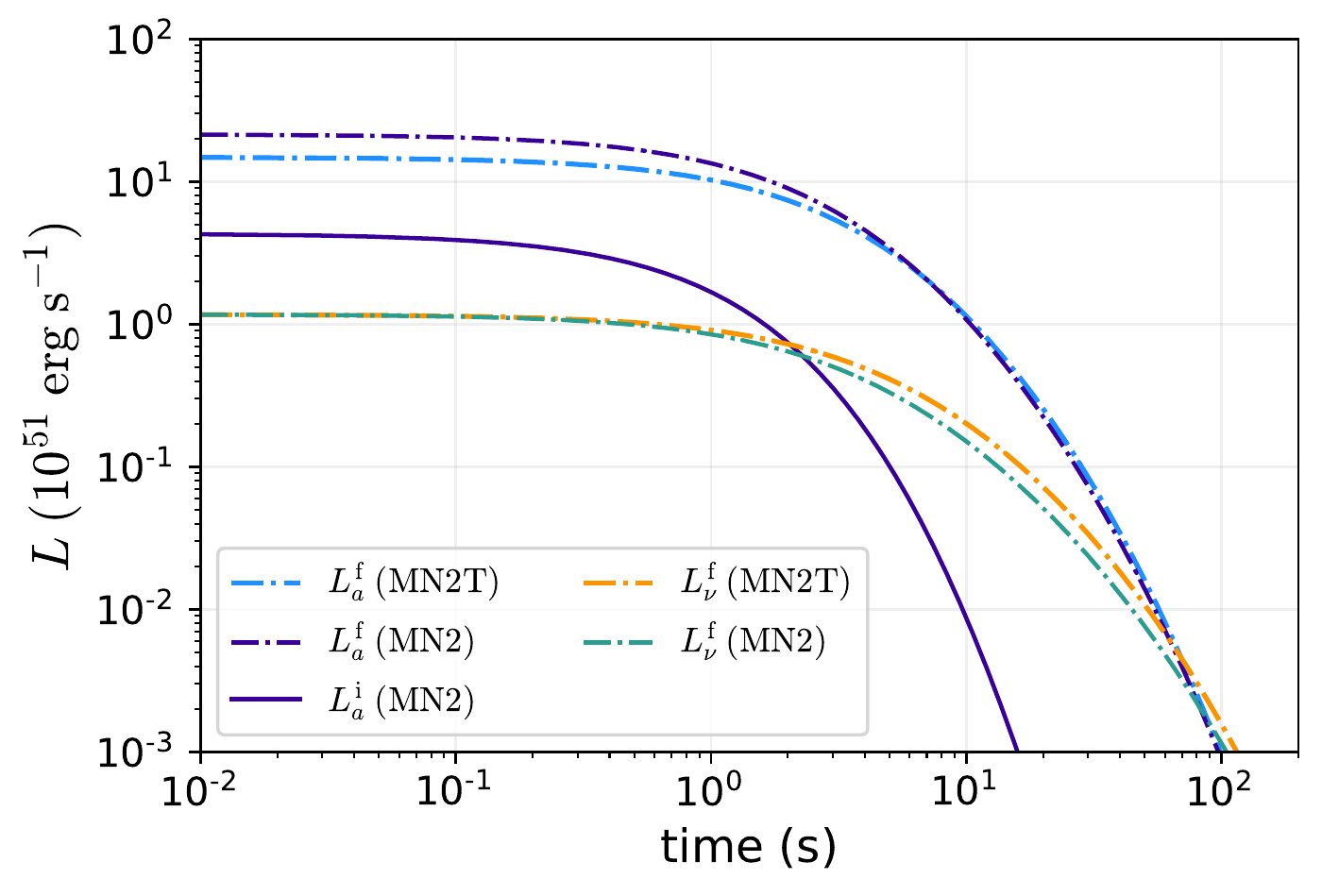}
%\end{center}
\caption{\label{fig:Ltemp} Comparison of the time evolution of neutrino and axion luminosities considering the original MN2 model and the modified MN2T with  temperature suppression. In dash-dotted dark (light) blue we have the final phase ALP luminosity for the MN2 (MN2T) model.  Analogously, the dash-dotted green (yellow) curves show the final phase neutrino luminosity for the MN2 (MN2T) model. The solid blue line indicates the initial phase contribution of the ALP luminosity.}
\end{figure}

Nevertheless, in contrast to the neutrino case, since ALP couplings are very suppressed, such particles should stream freely from the PNS center and, hence, their luminosity can also depend on the central temperatures. This implies that the final phase is relevant for the luminosity computation, and high temperature values can overestimate the total ALP luminosity. In order to estimate the impact of this issue on the ALP luminosity computation, we modified the 147S model to suppress the maximum temperature of the final phase down to values close to $\sim 40$ MeV (model 147ST). In Figure~\ref{fig:Ltemp} the dash-dotted blue curves show the comparison between the original final phase ALP luminosity (dark blue) and the one with the temperature suppression (light blue). Similarly, the green (yellow) dash-dotted curves show the final phase neutrino luminosity considering the 147S (147ST) model. At $t = 10$ s we estimate the difference in the axion (neutrino) luminosity to be of order of 5\% (25\%). For $t = 100$ s the impact increases to about 8\% (35\%). In the case of the Primakoff interaction, considering the model MP1 and employing the same temperature suppression, we estimate a difference of 38\% (0.7\%) in the axion (neutrino) luminosity for $t = 10$ s.

\vspace{-10pt}

%%%%%%%%%%%%%%%%%%%%%%%%%%%%%%%%%%%%%%%%%%%%%%%%%%%%%%%%%%%
\section{Outlook} \label{sec:outlook}
%%%%%%%%%%%%%%%%%%%%%%%%%%%%%%%%%%%%%%%%%%%%%%%%%%%%%%%%%%%

In the present work we introduced an analytic method to estimate the impact of the inclusion of axion-like-particle emission in supernovae explosions on neutrino observables, such as the luminosity $L_\nu$ and average energy $\langle E_\nu \rangle$. In particular, we considered the ALP emission via axion-nucleon Bremsstrahlung and axion-photon Primakoff interactions, and derived semi-analytic expressions for the respective axion luminosities $L_a$, given by equations~\eqref{eq:axBlum} and~\eqref{eq:axPlum}, following the same arguments discussed in Ref.~\cite{Suwa:2020nee}. With such expressions we could solve the first-order differential equation~\eqref{eq:difeq}, which relates the total SN binding energy loss rate with the neutrino and axion luminosities,  numerically to obtain the PNS entropy as a function of time $s(t)$. Afterwards, we  used $s(t)$ to compute the final expressions of $L_\nu$, $L_a$, and $\langle E_\nu \rangle$.

In order to solve the differential equation we need to fix six PNS free parameters (mass $M_{\rm PNS}$, radius $R_{\rm PNS}$, density correction factor $g$, opacity boosting factor $\beta$, total energy emitted by neutrinos $E_{\rm tot}$ and proton fraction $Y_p$) plus two axionic couplings (axion-nucleon coupling $g_{aNN}$ and axion-photon coupling $g_{a \gamma \gamma}$). With such choices we can analyze the impact on the neutrino luminosity and mean energy and study the deviations from the SM case. We found that, for the current allowed ALP parameter space, the axion nuclear Bremsstrahlung emission is dominant in comparison with the Primakoff interaction, which, in turn, is more relevant for later times in the PNS evolution. We showed that for axion-nucleon(-photon) couplings larger than $g_{aNN} \sim 10^{-11}$ ($g_{a \gamma \gamma} \sim 10^{-13} \, {\rm MeV}^{-1}$) the effect on the neutrino observables is larger than $5 \%$. We also analyzed the neutrino and ALP luminosity and energy profiles for different choices and combinations of axionic couplings.

We provide all our results and computations in the user-friendly python package \textsc{ARtiSANS}, which is publicly available on GitHub at \url{https://github.com/anafoguel/ARtiSANS} together with a Jupyter tutorial Notebook. The code allows the user to compute the neutrino luminosity and average energy as well as the ALP nuclear Bremsstrahlung and/or Primakoff emission $L_a$ for any choice of PNS and ALP input parameters. It can also compute the time evolution and the total binding energy carried by neutrinos and ALPs.

Let us mention that, although the axion nuclear Bremsstrahlung has been considered the main channel in the context of axion emission in the past years, it was recently suggested that the pion-induced reaction~\cite{Fischer:2021jfm,Carenza:2020cis} can be relevant as well. As was pointed out in Ref.~\cite{Fischer:2021jfm}, the pion Compton scattering predominates over the nuclear Bremsstrahlung for some PNS environment conditions, especially at high temperatures and mass densities. The effect of the inclusion of such process would be to increase the nuclear ALP emissivity and the neutrino luminosity deviation from the SM reference values.

Another important comment regarding the axion nuclear Bremsstrahlung computation is that we used the usual one-pion exchange (OPE) approximation, where the two-nucleon interaction is described via the exchange of one pion. As discussed in Ref.~\cite{Carenza:2019pxu}, one of the effects of going beyond this approximation is to decrease the axion luminosity.

Although the analytic calculations described in this work possess computational limitations when compared to complete SN simulations, they profit from other benefits, such as the need of few free parameters, which makes them simpler and faster. We have shown that such computations can be effectively employed to examine the main features and properties of the ALP and neutrino emissivities and mean energies for different model choices.

%%%%%%%%%%%%%%%%%%%%%%%%%%%%%%%%%%%%%%
\vspace{-5pt}
\begin{acknowledgments}
This work was partially supported by INCT-FNA (Process No. 464898/2014-5), CAPES (Finance Code 001), CNPq, and FAPERJ. A.L.F was supported by FAPESP under contract 2022/04263-5.
\end{acknowledgments}
%%%%%%%%%%%%%%%%%%%%%%%%%%%%%%%%%%%%%%

\appendix

\section{Details of the analytic computation} \label{app:difeq}
\vspace{-5pt}
In section~\ref{sec:analytic} we obtained the expression for the axion emission luminosity $L_a = L_a^{\rm brem} + L_a^{\rm prim}$ as a function of the PNS variables, the entropy $s$ and the dark axionic couplings. For a given choice of PNS parameters $(M_{\rm PNS}, R_{\rm PNS}, g \beta , Y_p)$ and dark couplings $(g_{aNN}, g_{a \gamma \gamma})$, we can solve the differential equation~\eqref{eq:difeq} to obtain the evolution of the entropy with time
\be \label{eq:difeq2}
 \frac{d s }{dt}= \qty(\frac{d E_{\rm th} }{ d s} )^{-1} (- 6 L_\nu - L_a )\,.
\ee
For instance, when considering only nuclear Bremsstrahlung, which is dominant for~$g_{a \gamma \gamma } \leq 10^{-13} \, {\rm MeV}^{-1}$, we end up with 
\begin{eqnarray}
 \frac{d \hat s }{dt} &=& - \Big[ \, 14.4 \times 10^{-3} \, (\hat M_{\rm PNS})^{-\frac{13}{15}}
 (\hat R_{\rm PNS})^{\frac45}  \qty(\frac{g \beta}{3})^{-\frac45} 
\hat s^{-\frac15}  \nonumber\\
&& \quad + \, \alpha_{aNN} \, 5.9 \times 10^{20} \, (\hat M_{\rm PNS})^{\frac{8}{3}} (\hat R_{\rm PNS})^{-8} \, \Big] \, {\rm s}^{-1} \nonumber \,,
\end{eqnarray}
where the hat indicates dimensionless variables, normalized to their respective typical values as considered in the equations of section~\ref{sec:analytic}. 

In order to solve the  differential equation above, we need to provide an initial condition for the entropy. For this purpose we use, as an approximation, the analytic expression for the entropy that solves the differential equation without the axionic contributions, i.e. \cite{Suwa:2020nee},
\be \label{eq:entropNu}
\hat s (t) = 4 \, (\hat M_{\rm PNS})^{\frac{13}{6}}  (\hat R_{\rm PNS})^{-2} \qty(\frac{g \beta}{3})^2 \qty(\frac{t + t_0}{100 \, \rm{s}})^{- \frac52} \, ,
\ee
where $t_0$ is the time origin, given by 
\be
t_0 = 210  \, \, (\hat M_{\rm PNS})^{\frac{6}{5}}  (\hat R_{\rm PNS})^{-\frac65} \qty(\frac{g \beta}{3})^{\frac45} (\hat E_{\rm tot})^{- \frac15} \, \rm{s} \, ,
\ee
in terms of the total energy emitted by neutrinos $\hat E_{\rm tot} \equiv E_{\rm tot}/(10^{52} \, {\rm erg})$. 

Hence, once we fix the set of PNS parameters $(M_{\rm PNS}, R_{\rm PNS}, g \beta ,  E_{\rm tot})$, we can use $t_0$ to obtain the initial entropy $\hat s (t_i) \equiv \hat s_i$
at a given initial time $t_i$. With this information we can numerically solve the differential equation~\eqref{eq:difeq2} to obtain the time evolution of the entropy for the chosen set of PNS and dark parameters. Finally, we can use the calculated entropy evolution to compute the neutrino and ALP luminosities, given by equations~\eqref{eq:nulum}, \eqref{eq:axBlum} and~\eqref{eq:axPlum}, and also the average energy of neutrinos~\cite{Suwa:2020nee}
\be
\langle E_\nu \rangle \simeq 7 \, (\hat M_{\rm PNS})^{\frac{1}{5}}  (\hat R_{\rm PNS})^{-\frac45} \qty(\frac{g \beta}{3})^{-\frac15} (\hat s)^{\frac35}  \MeV \, .
\ee

One last point important to highlight is that the computation of the differential equation should be divided into two phases. The reason is due to the fact that for early times in the PNS evolution there are no heavy nuclei in the crust~\cite{Suwa:2013mva}, which implies that the boosting factor $\beta$ acquires a time-dependence. Here, to simplify, we will follow~\cite{Suwa:2020nee} and consider an initial phase with small boosting ($\beta = 3$) and a later phase where the coherent scattering due to the heavier nuclei is relevant and, hence, the opacity boost increases ($\beta \gg 1$).

Since we divide the evolution in two components, the luminosities and neutrino mean energies will be expressed by
\begin{eqnarray}
L_{\nu,a} &=& L_{\nu,a}^i + L_{\nu,a}^f \, , 
\\ 
\langle E_\nu \rangle &=& \frac{L_{\nu}^i + L_{\nu}^f }{L_{\nu}^i/\langle E_\nu^i \rangle + L_{\nu}^f/\langle E_\nu^f \rangle } \,,
\end{eqnarray}
where the superscripts $({i,f})$ indicate the computation of the luminosity and average energy in the initial and final phases, respectively.
%%%%%%%%%%%%%%%%%%%%%%%%%%%%%%%%%%%%%%

%%%%%%%%%%%%%%%%%%%%%%%%%%
\bibliography{SNalps}

\end{document}